\documentclass[aps,prl,twocolumn,reprint,showpacs,superscriptaddress,nofootinbib]{revtex4}  
\usepackage{graphicx}  
\usepackage{dcolumn}   
\usepackage{bm}        
\usepackage{amssymb}   
\usepackage{setspace}
\usepackage{hyperref}
\usepackage{textcomp}
\usepackage{amsmath}
\usepackage{amsfonts}
\begin{document}

\title{Chaos Removal in the $R+qR^2$ gravity: the Mixmaster model}

\author{Riccardo Moriconi}
\email{moriconi@na.infn.it}
\affiliation{Dipartimento di Fisica, Universit\`a di Napoli ``Federico II'',
Compl. Univ. di Monte S. Angelo, Edificio G, Via Cinthia, I-80126, Napoli, Italy}
\affiliation{INFN Sezione di Napoli, Compl. Univ. di Monte S. Angelo, Edificio G, Via Cinthia, I-80126, Napoli, Italy}

\author{Giovanni Montani}
\email{giovanni.montani@frascati.enea.it}
\affiliation{Dipartimento di Fisica, Universit\`a degli studi di Roma "La Sapienza", P.le A. Moro 5 (00185) Roma, Italy}
\affiliation{ENEA, Unit\`a Tecnica Fusione, ENEA C. R. Frascati, via E. Fermi 45, 00044 Frascati (Roma), Italy}

\author{Salvatore Capozziello}
\email{capozziello@na.infn.it}
\affiliation{Dipartimento di Fisica, Universit\`a di Napoli ``Federico II'',
Compl. Univ. di Monte S. Angelo, Edificio G, Via Cinthia, I-80126, Napoli, Italy}
\affiliation{INFN Sezione di Napoli, Compl. Univ. di Monte S. Angelo, Edificio G, Via Cinthia, I-80126, Napoli, Italy}
\affiliation{Gran Sasso Science Institute (INFN), Via F. Crispi 7, I-67100, L' Aquila, Italy}
\date{\today}

\begin{abstract}
We study the asymptotic dynamics
of the Mixmaster Universe, near the cosmological
singularity, considering $f(R)$ gravity up to a quadratic corrections in the Ricci scalar $R$.
The analysis is performed in the scalar-tensor framework
and adopting Misner-Chitr\'e-like variables to
describe the Mixmaster Universe, whose dynamics
resembles asymptotically a billiard-ball in a
given domain of the half-Poincar\'e space.
The form of the potential well depends on the spatial
curvature of the model and on the particular form
of the self-interacting scalar field potential.
We demonstrate that the potential walls determine
an open domain in the configuration region,
allowing the point-Universe to reach the absolute
of the considered Lobachevsky space. In other words,
we outline the existence of a stable final Kasner regime
in the Mixmaster evolution, implying the
chaos removal near the cosmological singularity.
The relevance of the present issue relies both on the
general nature of the considered dynamics,
allowing its direct extension to the BKL conjecture too,
as well as the possibility to regard the considered
modified theory of gravity as the first correction to
the Einstein-Hilbert action as a Taylor expansion
of a generic function $f(R)$ (as soon as a cut-off on the space-time curvature
takes place).
\end{abstract}

\maketitle
\section*{Introduction}
The chaotic dynamics of the Mixmaster Universe
\cite{BKLref1},\cite{mixmastermisner},\cite{misner} is a basic prototype of the local
(sub-horizon) behaviour of the generic cosmological
solution (the so-called BKL conjecture\cite{BKLref2}).
Investigating the stability of such a chaotic picture with respect to the presence of matter \cite{scalar field},\cite{montanireview},\cite{primordial}
and space-time dimensions number\cite{hennaux1},\cite{hennaux2},\cite{del1}
 has seen a great effort over the last four decades
 and the most significant issue was the proof of
 the chaos removal when a massless scalar field
 is involved in the dynamics \cite{scalar field}.
 Such a result is a consequence of the capability
manifested by the scalar field kinetic energy of affecting
the second (quadratic) Kasner condition, easily restated in the Hamiltonian picture, as shown in \cite{berger}.
This property of the massless scalar field acquires
intriguing perspectives when $f(R)$ modified theory
of gravity are considered \cite{capfirst},\cite{capozzielloreview},\cite{odintsovreview},\cite{odintsov},\cite{guo}. In fact, these alternative
formulation of the gravitational field dynamics can
be represented by an equivalent scalar-tensor picture: the scalar degree of freedom associated to
the form of the function $f$ is expressed via a
self-interacting scalar field, coupled to the ordinary
General Relativity \cite{bar1},\cite{bar2},\cite{bar3},\cite{del2}.
When implementing this scalar-tensor scheme to the
Mixmaster Universe dynamics, a natural question arise:
the kinetic term of the scalar field removes
the chaotic behaviour, but the presence of a
potential term could restore it? Thus we can study, for specific modified theories of
gravity, if the Mixmaster chaos
survives or not, simply characterizing the corresponding
scalar field potential.
Here we analyse the modified gravity theory
corresponding to a quadratic correction in the Ricci scalar
to the ordinary Einstein-Hilbert Lagrangian,
both because it is the simplest viable deviation
from General Relativity (apart from a cosmological
constant term), as well as the first correction emerging
from a Taylor expansion of a $f(R)$ theory for
very small values of the space-time Ricci scalar,
i.e. for very law curvatures, like we observe today
in the Solar System\cite{staro1}.
The quadratic term in the Ricci scalar provides
an exponential-like potential term for the
self-interacting scalar field, when a scalar-tensor
reformulation of the model is considered.
This case is particularly appropriate to the analysis
we pursue of the Mixmaster dynamics 
in terms of the Misner-Chitr\'e-like variables \cite{primordial},\cite{montani-kirillov},\cite{imponente-montani},\cite{MCthesis}.
In fact, the kinetic term of the scalar field is
on the same footing of the anisotropy term contribution
and, for the considered Lagrangian, also the potential
term is isomorphic to the spatial curvature of
the model, i.e. the total potential term is constituted by equivalent exponential profile.
In the asymptotic limit toward the initial singularity
the total potential takes the form of four potential walls,
whose morphology determines if the configuration domain is closed or not.
Indeed, we demonstrate how the whole domain,
available in principle, is a constant negative curvature space (half-Poincar\`e space).
We first analyse the case of
the Mixmaster Universe in the presence of a massless
scalar field, demonstrating the open nature of its
configuration space and the implied existence of a
stable Kasner regime to the initial singularity.
Then, we face a detailed study of the dynamics in the
presence of the total potential and the still open
structure of the configuration domain.
Thus, we demonstrate the non-chaotic nature
of the Mixmaster Universe behaviour, as it is described
by the scalar-tensor version of the $R^2$-gravity.
\section*{$f(R)$ gravity}
The $f(R)$ theories of gravity are a direct generalization of the Einstein-Hilbert Lagrangian
consisting in a replacement of the Ricci Scalar $R$ by a general function $f(R)$\cite{capozzielloreview},\cite{odintsovreview},\cite{odintsov},\cite{odintsov2}:
\begin{equation}
\label{S}
S=\frac{1}{16\pi}\int d^{4}x \sqrt{-g}f(R)
\end{equation}
where $g$ is the determinant of the metric\footnote{We use the $(-,+,+,+)$ signature of the metric and the geometric unit system $(c=G=\hbar=1)$.}.
The introduction of the additional degree of freedom, related to the presence of the $f(R)$ term, can be translated into a dynamics of a self-interacting scalar field coupled with the Einstein-Hilbert Action, the so-called \textit{Scalar-Tensor framework}.
In this approach,  a new auxiliary field $\chi$ is introduced to get the following equivalent version of the action (\ref{S}):
\begin{equation}
\label{ausi}
S=\frac{1}{16\pi}\int d^{4}x \sqrt{-g}[f(\chi) - f'(\chi)(R-\chi)].
\end{equation}
The variation of the action (\ref{ausi}) with respect to $\chi$ provides $f''(\chi)(R-\chi)=0$, implying $\chi=R$ if $f''(\chi) \neq 0$. By a redefinition of the auxiliary field $\chi$ in the form $\varphi=f'(\chi)$ the action becomes
\begin{equation}
\label{ausi-2}
 S=\frac{1}{16\pi}\int d^{4}x \sqrt{-g}[\varphi R - \chi(\varphi)\varphi + f(\chi(\varphi))].
\end{equation}
It is now possible to perform a conformal transformation on the metric $g_{\mu \nu} \rightarrow \tilde{g_{\mu \nu}} =  \varphi g_{\mu \nu}$ and a scalar field redefinition $\varphi \equiv f'(R) \rightarrow \phi = \sqrt{\frac{3}{16\pi}}\ln f'(R)$ in order to obtain
\begin{equation}
\label{azione}
S=\int d^{4}x \sqrt{-\tilde{g}}\left[ \frac{\tilde{R}}{16\pi} - \frac{1}{2}\partial^{\alpha}\phi\partial_{\alpha}\phi - U(\phi) \right],
\end{equation}
where the potential term $U(\phi)$ has the form:
\begin{equation}
\label{pote-cam-sca}
U(\phi) = \frac{Rf'(R)-f(R)}{16 \pi (f'(R))^{2}}.
\end{equation}
For small values of the Ricci scalar, the first order correction to the Einstein-Hilbert Lagrangian,  is represented by a quadratic correction, i.e.
\begin{equation}
\label{f}
f(R) = R+q R^2.
\end{equation}
By this choice, the potential term (\ref{pote-cam-sca}) takes the form
\begin{equation}
\label{potcamsca}
U(\phi) = \frac{1}{64\pi q}\left( 1 - 2\exp^{-4\sqrt{\frac{\pi}{3}}\phi}+\exp^{-8\sqrt{\frac{\pi}{3}}\phi} \right).
\end{equation}
This is the effective potential that emerges in the so called Starobinsky-inflation model\cite{staro1}. Such a model ensures a "slow-rolling" face and it is an inflationary model passing the latest inflation constraint\cite{planck}.
\section*{The Mixmaster model dynamics}
Following the standard representation of the Bianchi IX model\cite{gravitation} in the Misner variables\cite{mixmastermisner},\cite{misner} the Einstein-Hilbert action takes the form:
\begin{equation}
\label{azioneMisner}
S_{g} = \int dt \left[ p_{\alpha}\dot{\alpha} + p_{+}\dot{\beta_{+}} + p_{-}\dot{\beta_{-}} - \frac{Ne^{-3\alpha}}{24\pi}\mathcal{H}_{IX} \right],
\end{equation}
where the dynamics of the model implies the superHamiltonian constraint
\begin{equation}
\label{vincolo misner}
\mathcal{H}_{IX} \equiv -p_{\alpha}^{2}+p_{+}^{2}+p_{-}^{2}+12 \pi^{2} e^{4\alpha}V_{IX}(\beta _{\pm}) = 0.
\end{equation}
Here $\alpha$ expresses the isotropic component of the Universe(i.e. the volume of the universe) and the initial singularity is reached for $\alpha\rightarrow -\infty$, while the traceless matrix $\beta_{ab} = diag(\beta _{+} + \sqrt{3}\beta _{-} , \beta _{+} - \sqrt{3}\beta _{-} , -2\beta _{+})$ accounts for the anisotropy of this model.
Furthermore, $p_{\alpha},p_{\pm}$ are the conjugated momenta to $\alpha,\beta_{\pm}$ respectively and $V_{IX}(\beta _{\pm})$ is the potential term depending only on $\beta_{\pm}$, corresponding to the spatial curvature. If we execute an ADM reduction of the dynamics\cite{ADM}, the Bianchi IX model resembles the behaviour of a two-dimensional particle, evolving with respect to the time-like variable $\alpha$ in the $\beta_{+},\beta_{-}$ plane. By other words, the system dynamics is summarized by the time-dependent hamiltonian $H_{IX}$:
\begin{equation}
\label{HBIXmisner}
-p_{\alpha} = H_{IX} \equiv \sqrt{p_{+}^{2}+p_{-}^{2}+12 \pi^{2} e^{4\alpha}V_{IX}(\beta _{\pm})}.
\end{equation}
Looking at the form of the potential term $V_{IX}(\beta _{\pm})$, it is possible, taking into account the three leading terms, to parametrize it as an infinitely steep potential well\cite{misner}. This way, the point-Universe lives inside the triangular region of the configuration space where the potential term is negligible; such a region it is individuate when the following three conditions hold:
\begin{equation}
\begin{split}
\label{anisotropyMisner}
&\frac{1}{3} + \frac{\beta_{+}+\sqrt{3}\beta_{-}}{3\alpha} > 0,\\
&\frac{1}{3} + \frac{\beta_{+}-\sqrt{3}\beta_{-}}{3\alpha} > 0,\\
&\frac{1}{3} - \frac{2\beta_{+}}{3\alpha} > 0.
\end{split}
\end{equation}
The presence of the ``time'' variable $\alpha$ in the relations (\ref{anisotropyMisner}) causes the outside motion of the potential walls and the corresponding time-dependence of the domain allowed to the point-Universe motion.
Such a dependence can be removed in the framework of the \textit{Misner-Cithr\`e} variables  $\tau,\zeta,\theta$\cite{gravitation},\cite{MCthesis} as standing in the \textit{Poincare Half-Plane}:
\begin{equation}
\begin{split}
\label{MCtrans}
&\alpha - \alpha_{0} = -e ^{\tau}\frac{1+u+u^2 + v^2}{\sqrt{3}v},\\
&\beta_{+} = e^{\tau}\frac{-1+2u+2u^2+2v^2}{2\sqrt{3}v}\\
&\beta_{-} = e^{\tau}\frac{-1-2u}{2v}.
\end{split},
\end{equation} 
where $-\infty < \tau < \infty$, $-\infty < u < +\infty$, $0 < v < +\infty$. In this scheme the role of the hamiltonian time is assigned to $\tau$ and the singularity is approach for $\tau\rightarrow\infty$.
The transformations (\ref{MCtrans}) permit to rewrite the conditions (\ref{anisotropyMisner}) as independent of the variable $\tau$ and thus the domain within which the particle lives is fixed respect to the time variable.
Making use of the transformations (\ref{MCtrans}), the  Hamiltonian (\ref{HBIXmisner} in the free-potential case rewrites as
\begin{equation}
\label{vincolo MC}
-p_{\tau} = H_{IX} \equiv  \sqrt{v^2(p_{u}^{2} + p_{v}^{2} ) },
\end{equation}
and the point-Universe lives in the $u,v$ plane inside the region individuate when the following three conditions hold:
\begin{equation}
\begin{split}
\label{anisotropyUV}
& \frac{-u}{1+u+u^2+v^2} >0, \\
&\frac{1+u}{1+u+u^2+v^2} > 0,\\
&\frac{u(u+1) + v^2}{1+u+u^2+v^2}>0.
\end{split}
\end{equation}
As shown by \cite{imponente-montani}, the asymptotic evolution towards the singularity is covariantly chaotic because it is isomorphic to a billiard on the Lobachevsky plane. This demonstration is based on three points:i)the Jacobi metric in the $u,v$ plane has a negative constant curvature; ii)the Lyapunov exponent, defined as in \cite{pesin}, are greater than zero; iii)the configuration space is (dinamically) compact. The occurrence of the these three properties ensures that the geodesic trajectories cover the whole configuration space.
\\
\section*{Mixmaster Universe in the $R^{2}$-gravity}
\begin{figure}[h!]
\centering
\includegraphics[scale=.69]{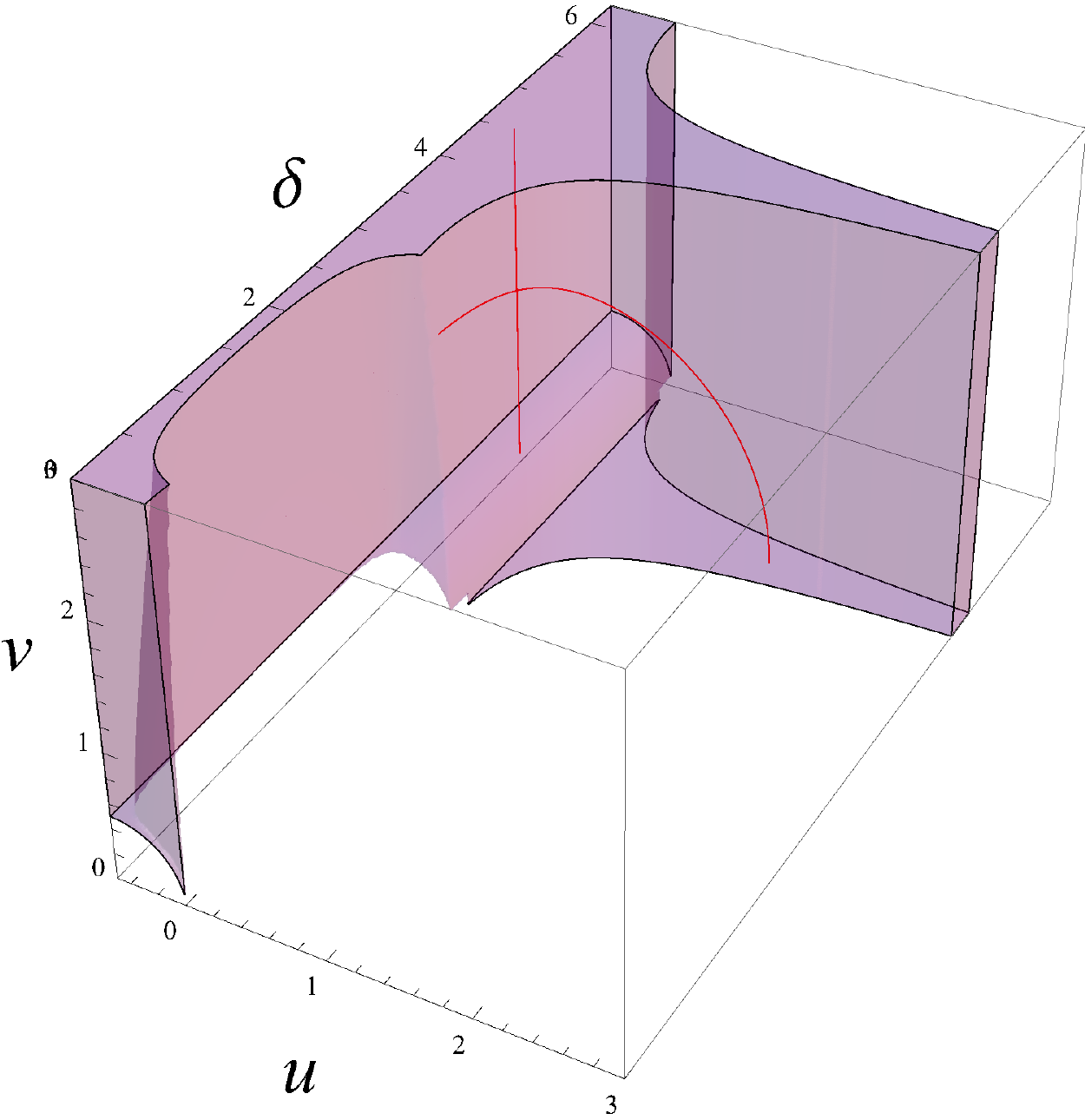}
\caption{ \footnotesize The point-Universe lives inside the region marked by the walls, where the conditions (\ref{anisotropyUVdeltacompleto}) are verified. We also sketch the trajectories reaching the absolute.}
\label{fig:potBIXUVphi-scalare}
\end{figure}

Now we analyse the case of the gravitational Lagrangian (\ref{f})
when the Bianchi IX model is considered.
As starting point we consider the modified gravity model (\ref{azione}) in terms of the variables $\alpha,\beta_{+},\beta_{-},\phi$. Following the same procedure of the previous section we get the generalized reduced hamiltonian $-p_{\alpha} = H$ of the form
\begin{equation}
\label{ridottamisner}
H \equiv \sqrt{p_{+}^{2}+p_{-}^{2}+p_{\phi}^{2}+ 12 \pi^{2} e^{4\alpha}V_{IX} + 4e^{6\alpha}U}.
\end{equation}
In Eq.(\ref{ridottamisner}), we rescaled the zero point of $\alpha\rightarrow \alpha - \alpha_{0}$, so that the spatial metric factor $e^{3\alpha} \rightarrow \frac{1}{(6\pi)}e^{3\alpha}$, and a redefinition of the scalar field amplitude $\phi \rightarrow \sqrt{2}(6\pi)\phi$ is considered too.
A natural parametrization, in the \textit{Misner-Cithr\`e - Poincare Half-Plane} scheme, that reduces to the relations (\ref{MCtrans}) if the scalar field is turned-off, reads as follows
\begin{equation}
\begin{split}
\label{MCphitrans}
&\alpha - \alpha_{0} = -e ^{\tau}\frac{1+u+u^2 + v^2}{\sqrt{3}v},\\
&\beta_{+} = e^{\tau}\frac{-1+2u+2u^2+2v^2}{2\sqrt{3}v},\\
&\beta_{-} = e^{\tau}\frac{-1-2u}{2v} \cos \delta,\\
&\phi =  e^{\tau}\frac{-1-2u}{2v}  \sin \delta, \\
\end{split}
\end{equation} 
where $-\infty<\tau<\infty$, $-\infty < u < +\infty$ , $0 < v < +\infty$ and $0<\delta<2\pi$.
In this new system of variables the reduced Hamiltonian takes the form:
\begin{equation}
\label{vinUV}
-p_{\tau} = H \equiv  \sqrt{ v^2\left[p_{u}^{2} + p_{v}^{2}+4\frac{p_{\delta}^{2}}{(1+2u)^2}\right] + e^{2\tau}\mathcal{V}}.
\end{equation}
The introduction of the degree of freedom related to the scalar field implies that the point-Universe lives inside a 3-dimensional domain in the configuration space $u,v,\delta$ determined by the potential term:
\begin{widetext}
\begin{multline}
\label{potenzialeintero}
e^{2\tau}\mathcal{V} = e^{2\tau}[ 12\pi^{2}e^{-4e^{\tau}\xi(u,v)}V_{IX}(u,v,\delta,\tau) + 4e^{-6e^{\tau}\xi(u,v)}U(u,v,\delta,\tau)] = \\ =  12\pi^{2}e^{2\tau}\left( e^{-\frac{12e^{\tau}}{\sqrt{3}v}(u+u^{2}+v^{2})} + e^{-\frac{6e^{\tau}}{\sqrt{3}v}(1+(1+2u)\cos \delta)} + e^{-\frac{6e^{\tau}}{\sqrt{3}v}(1-(1+2u)\cos \delta)} \right) + \\ + \frac{e^{2\tau}}{8\pi q}\left( e^{-\frac{12e^{\tau}}{\sqrt{3}v}(1+u+u^{2}+v^{2})} - 2e^{-\frac{6e^{\tau}}{\sqrt{3}v}(1+u+u^{2}+v^{2} -2\sqrt{2\pi^{3}}(1+2u)\sin \delta)} + e^{-\frac{6e^{\tau}}{\sqrt{3}v}(1+u+u^{2}+v^{2} -4\sqrt{2\pi^{3}}(1+2u)\sin \delta)} \right),
\end{multline}
\end{widetext}
where $\xi(u,v) = \frac{1+u+u^2 + v^2}{\sqrt{3}v}$. Due to the exponential forms of the terms in Eq.(\ref{potenzialeintero}), when the singularity is approached ($\tau\rightarrow\infty$) the point-Universe is confined to live inside a 3-dimensional domain defined as the region where all the exponents of the six terms are simultaneously greater than zero.
Looking the Eq.(\ref{potenzialeintero}), the potential term $\mathcal{V}$ behaves as an infinitely steep potential well as in the Poincar\`e variables (\ref{anisotropyUV}). So for the evolution of the point-Universe it is possible to neglect the potential everywhere  in a suitable domain. 
As first step we study the case in absence of all the potential terms ($\mathcal{V}=0$), i.e. we deal with the Hamiltonian problem: 
\begin{equation}
H = v\sqrt{p_{u}^{2} + p_{v}^{2}+4\frac{p_{\delta}^{2}}{(1+2u)^2}}.
\end{equation}
The hamiltonian equations for this potential-free system (Bianchi I model with the massless scalar field) are
\begin{equation}
\begin{split}
\label{MCphitrans}
&\dot{u} = \frac{\partial H}{\partial p_{u}} = \frac{v^2}{\epsilon}p_{u} \quad , \quad \dot{p_{u}} = -\frac{\partial H}{\partial u} = \frac{8v^2}{\epsilon}\frac{p_{\delta}^2}{(1+2u)^3}\\
&\dot{v} = \frac{\partial H}{\partial p_{v}} = \frac{v^2}{\epsilon}p_{v} \quad , \quad  \dot{p_{v}} = -\frac{\partial H}{\partial v} = -\frac{\epsilon}{v}\\
&\dot{\delta} = \frac{\partial H}{\partial p_{\delta}} = \frac{4v^2}{\epsilon}\frac{p_{\delta}}{(1+2u)^2} \quad , \quad \dot{p_{\delta}} = -\frac{\partial H}{\partial \delta}=0.
\end{split}
\end{equation} 
It is possible to demonstrate, as we approach the singularity,  that $H$ is a constant of motion with respect to the ``time'' variable $\tau$, following \cite{imponente-montani}. Thus, we perform the substitution $H\simeq \epsilon=const.$ inside Eq's(\ref{MCphitrans}).
It is now possible, by following the Jacobi procedure\cite{arnold} and using the equations of motion (\ref{MCphitrans}), to write down the line element for the three-dimensional Jacobi metric in terms of the configuration variables, i.e.	
\begin{equation}
\label{elementolinea}
ds^{2} = \frac{\epsilon}{v^2}\left[ du^{2} + dv^{2} + \frac{(1+2u)^{2}}{4}d\delta^{2} \right].
\end{equation}
By a direct calculation we see that this metric has a negative constant curvature (the associated Ricci scalar is $R=-\frac{6}{\epsilon}$) and then the point-Universe moves over a negatively curved three-dimensional space.
Furthermore, we can find two singular values for the metric in correspondence to $u=-\frac{1}{2}$, $v = 0$. 
This feature allows us to restrict the domain of the configuration space in which we will study the trajectories of the point-Universe to the fundamental one identified by the inequalities $-\frac{1}{2} < u < +\infty$, $0 < v < +\infty$, $0 < \delta < 2\pi$. Indeed there is no way for the point-Universe trajectories to cross over the two planes $u=-\frac{1}{2}$, $v=0$ (each choice of the Lobachevsky ``half-space'' is equivalent respect to the other one).
The intermediate step toward the general case of the potential (\ref{potenzialeintero}), corresponding to the ordinary Mixmaster model in the presence of a massless scalar field, takes place when we retain only the exponential terms due to the spatial curvature, namely $\mathcal{V}\simeq 12\pi^{2}e^{-4e^{\tau}\xi(u,v)}V_{IX}(u,v,\delta,\tau)$. Then, the point-Universe lives in the region where are simultaneously satisfied the three following conditions
\begin{equation}
\label{anisotropyUVdelta}
\begin{cases} 
1+(1+2u)\cos \delta > 0, \\ 1-(1+2u)\cos \delta > 0, \\ u(u+1) + v^2 >0. 
\end{cases}
\end{equation}
\begin{figure}[h!]
\centering
\includegraphics[scale=.7]{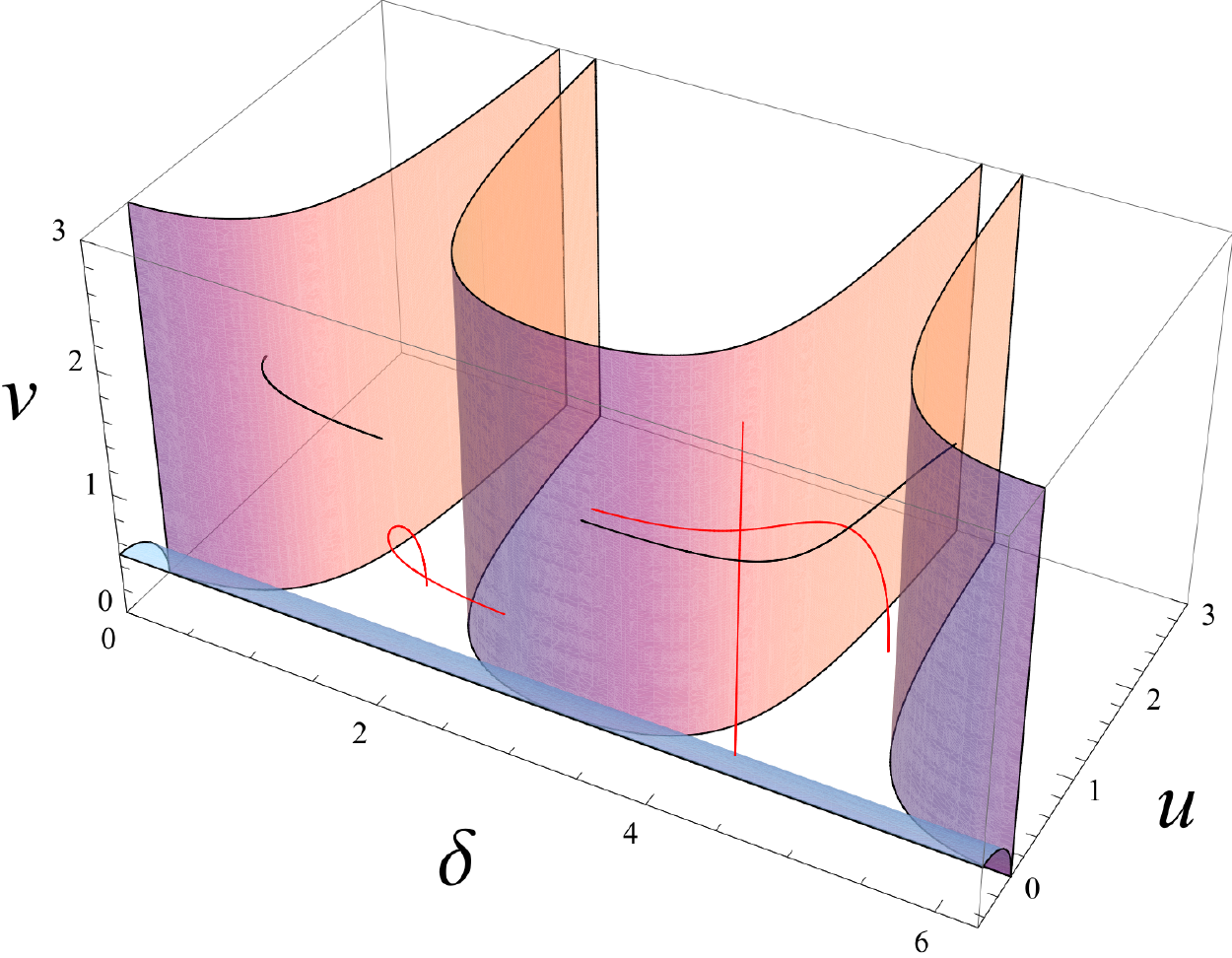}
\caption{ \footnotesize The black lines represent the trajectories associated to a points-Universe that bounce against the walls. Instead, the red lines describe the points-Universe witch directly approach the singularity.}
\label{fig:traiettorie}
\end{figure} 
We now implement a numerical integration of the system (\ref{MCphitrans}) in order to analyse the behaviour of the trajectories in the potential free region and then use this result for interpreting the effect of the scalar curvature. As we can see in the Fig.\ref{fig:traiettorie} an opening of the domain emerges due to the presence of the scalar field and it is possible to individuate two families of trajectories: those ones corresponding to a point-Universe that bounces against the walls and turn back inside the domain (the black ones) and those corresponding to a particle that approach the so called ``absolute''\cite{kirillov1997} (the red ones), for values $v\rightarrow0,\infty$, with no other bounces until the singularity.
The presence of the trajectories of the second family shows the removal of the oscillatory behavior of the Mixmaster model coupled with a massless scalar field \cite{BKLref2},\cite{berger}.
Let us see what happen if we consider the complete potential term(\ref{potenzialeintero}). This time the restrictions on the dynamics imply that the particle is confined inside a region where all the six exponential terms in Eq.(\ref{potenzialeintero}) are simultaneously greater than zero. We can immediately remove one of the six conditions because the first exponent related to the potential of the scalar field $1+u+u^2+v^2$ is always greater than zero for any values of $u,v$ taking in consideration. Thus, the five conditions that identify the domain are
\begin{equation}
\label{anisotropyUVdeltacompleto}
\begin{cases} 
1+(1+2u)\cos \delta > 0 \\ 1-(1+2u)\cos \delta > 0 \\ u(u+1) + v^2 >0 \\ 1+u+u^{2}+v^{2} -2\sqrt{2\pi^{3}}(1+2u)\sin \delta >0 \\ 1+u+u^{2}+v^{2} -4\sqrt{2\pi^{3}}(1+2u)\sin \delta >0
\end{cases}
\end{equation}
We observe that the last of the conditions above naturally implies the validity of the fourth one too. Thus, we indeed deal with four potential walls only.
As we can see in the Fig.\ref{fig:potBIXUVphi-scalare}, taking into account also the potential term $U(u,v,\delta,\tau)$ implies that the available configuration space for the point-Universe is clearly reduced with respect to the case $U=0$ (see Fig.\ref{fig:traiettorie}). 
However, trajectories yet exist(the red lines in the Fig.\ref{fig:potBIXUVphi-scalare}) corresponding to a point-Universe that is able to reach the absolute for $v\rightarrow0,\infty$. 
For this reason we can firmly conclude that a quadratic correction in the Ricci scalar to the Einstein-Hilbert action, that in the Scalar-Tensor theory is equivalent to the dynamics  of a self-interacting scalar field (with potential terms of the form (\ref{potcamsca})), is able to remove the never-ending bounces of the point-Universe against the walls. As a result of the bounces against the infinite potential walls (which can be described by a reflection rule\cite{montani-kirillov},\cite{indicikasner}), soon or later the point-Universe reach a trajectory connected with the absolute. It is worth noting that the analysis above 
is referred to the choice $q>0$, in which case the sign of the scalar field potential is the same one of the scalar curvature. This choice is forced by the request that the additional scalar mode, associated to the quadratic modification, be a real (non-tachyonic) massive one, 
accordingly to the original Starobinsky approach in \cite{staro1} and demonstrated also in  \cite{segnoq}.
However, in the case $q<0$, the scalar field  potential would not contribute an infinite positive wall, but an infinite depression. Since in the region of zero potential, the point-Universe has always positive ``energy'', we can easily conclude that such a case overlaps the non-chaotic potential-free one. We now observe that, in correspondence to 
the configuration region $v=0,\infty$ and  
$\delta = \frac{3\pi}{2}$, the scalar field 
acquires negative diverging values 
and its potential terms manifests a 
diverging behaviour. Such a profile of the 
scalar field is typical of a Bianchi I 
solution near the singularity \cite{scalar field} and the diverging character 
of the potential term means that 
General Relativity can not be asymptotically recovered. Rigorously speaking,
the 
present result on the chaos structure
applies to a quadratic correction in 
the Ricci scalar only, because it is the first terms of the Taylor expansion of the function $f(R)$ working nearby the singularity.
Nonetheless, our analysis has a general 
validity, as soon as, we take into 
account a physical cut-off at the Planck 
time, where classical theory starts to 
fail and a quantum treatment is required.
In fact, the Planckian cut-off would 
remove the $\phi$ and $U(\phi)$ divergences, 
allowing the Taylor expansion for 
$q \lesssim \frac{(ct_{cut})^2}{l_P^2}$, 
where $t_{cut}$ being the cut-off time 
and $l_P$ the Planck length. 
Since $t_{cut} > \frac{l_p}{c}$, we deal 
with the (non-severe) restriction 
$q\lesssim1$ for preserving the general nature 
of our result. This estimation follows requiring 
$R > qR^2$ and remembering that for the 
case of a Kasner solution, in the presence of 
a potential-free scalar field, 
the Ricci scalar behaves as $R\sim \frac{1}{t_{s}^2}$, where $t_{s}$ is the synchronous time.
We stress that qualitatively, a similar 
argument is at the ground of the non-chaotic nature of the Bianchi IX Loop 
Quantum dynamics in the semi-classical 
limit \cite{boj}.
However, the field $\phi(t_{s})$ admits, both for $v\rightarrow 0, \infty$ and $\delta = \pi/2$, trajectories implying its positive divergence. For such behaviours, corresponding to an open region in the initial condition, the potential $U(\phi)$ approaches a constant value and $\phi$ is effectively massless. It is just the existence of these diverging profiles at the ground of the chaos removal in the present model. The massless nature of the potential along specific trajectories is a good criterion for determining the chaotic properties of the Mixmaster Universe in a specific non-expanded $f(R)$ model. In fact, The behavior of the free scalar field 
reads $\phi_{f}(t_{s}) \propto \ln t_{s}$ and the 
corresponding kinetic energy density 
stands as $1/t_{s}^2$. Then, for a given 
$f(R)$ model, fixing the potential 
$U(\phi)$, the chaos removal is 
ensured by the validity of the condition $\lim _{t_{s} \rightarrow 0} U(\phi_{f}(t_{s}))t_{s}^2 = 0
$.
Clearly, the non-chaoticity is ensured 
if such a limit holds for a non-zero 
measure set of trajectories.
\section*{Conclusions}
The analysis above demonstrated how
including a quadratic correction in
the Ricci scalar to the Einstein-Hilbert
Lagrangian of the gravitational field
gives a deep insight on the nature of
the Mixmaster singularity: the evolution
of the scale factors is no longer chaotic
and a stable Kasner regime emerges as
the final approach to the
singular point. 

The relevance of this result is
in its generality with respect to the behaviour of the cosmological gravitational field. In fact, on one hand,
the result we derived in the homogeneous
cosmological setting, can be naturally extended
to a generic inhomogeneous Universe,
simply following the line of investigation
discussed in \cite{BKLref2},\cite{scalar field}.
 
The basic statement, at the ground of the BKL conjecture, is the space point decoupling 
in the asymptotic dynamics toward the 
cosmological singularity.  Such a dynamical 
property of a generic inhomogeneous cosmological 
model allows to reduce the behavior of a 
sub-horizon spatial region \cite{BKLref2},\cite{M95} 
to the prototype offered by the homogeneous 
Mixmaster Universe. We are actually stating that 
the time derivative of the dynamical variables 
asymptotically dominate their spatial gradients, 
limiting the presence of the spatial coordinates 
in the Einstein equation to a pure parametrical 
role. We are speaking of a conjecture because 
the chaotic features of the point-like dynamics 
induce a corresponding stochastic behaviour of 
the spatial dependence and the statement above 
requires a non-trivial treatment for its proof. 
Nonetheless a valuable estimation of the spatial 
gradient behaviour, when the space-time takes 
the morphology of a foam, is provided in 
\cite{K93}.
When a scalar field is present the situation is even more simple, because, after a certain number of iterations of the BKL map, in each space point, a stable Kasner regime takes place \cite{K87} and the validity of the solution is rigorously determined \cite{LK63}.  
Thus, we can extend our result to a generic 
inhomogeneous cosmological model simply 
considering the dynamical variables as 
space-time functions $u = u(\tau ,x^i)$, $v=v(\tau ,x^i)$ and $\delta = \delta (\tau ,x^i)$, which, in each space point, live in a half-Poincar  space and are governed by an independent and morphologically equivalent dynamics.
On the other hand,
the extension of General Relativity
we considered here is the most simple and
natural one, widely studied in literatures
in view of its implications on the
primordial Universe features.
Since the classical evolution is expected
to be predictive up to a finite value of
the Universe volume, i.e. up to a
given amplitude of the space-time curvature,
for sufficiently small coupling constant
$q$ values, the present model can be considered as
the quadratic Taylor expansion of a generic
$f(R)$ theory and we can then guess that
the non-chaotic feature is a very general
dynamical property, at least within the classical
domain of validity of the $f(R)$ theory.
In this sense we traced a very general and
reliable properties of the cosmological
gravitational field in modified theories of
gravity of significant impact on the
so-called billiard representation of the
generic primordial Universe\cite{hennaux1},\cite{montani-kirillov},\cite{indicikasner}.


\addcontentsline{toc}{chapter}{Bibliografia}
\begin{thebibliography}{9}

\bibitem{BKLref1}
V.A. Belinskii, I.M. Khalatnikov, E.M. Lifshttz, \textit{Oscillatory approach to a singular point in the relativistic cosmology}, Adv. Physics, 19(80), 525-573 (1970).

\bibitem{mixmastermisner}
C. W. Misner, \textit{Mixmaster Universe}, Phys. Rev. Letters {\bf 22}, 1071, 1969).

\bibitem{misner}
C. W. Misner, \textit{Quantum Cosmology I}, (PhysRev. {\bf 186}, 1319, 1969).

\bibitem{BKLref2}
V. A. Belinski, I. M. Khalatnikov and E. M. Lifshitz, \textit{A General Solution of the Einstein Equations with a Time Singularity}, (Adv.Phys. {\bf 31}, 639, 1982).

\bibitem{scalar field}
V. A. Belinskii, I. M. Khalatnikov, \textit{Effect of scalar and vector fields on the nature of the cosmological singularity}, (Sov. Phys. JETP 36(4), 591-597 , 1973).

\bibitem{montanireview}
G. Montani, M. V. Battisti, R. Benini, G. Imponente,
\textit{Classical and quantum features of the Mixmaster singularity}, (International
Journal of Modern Physics A {\bf 23}, pp. 23532503, 2008).

\bibitem{primordial}
G. Montani, M. V. Battisti, G. Imponente, R. Benini,
Primordial cosmology, (World Scientific, 2011).

\bibitem{hennaux1}
J. Demaret, M. Henneaux and P. Spindel, Phys. Lett., 164B, 27, (1985).

\bibitem{hennaux2}
Y. Elskens and M. Henneaux, Nucl. Phys., B290, 111 (1987).

\bibitem{del1}
N. Deruelle and P. Spindel, Classical Quantum Gravity 7, 1599 (1990).



\bibitem{berger}
B. K. Berger, \textit{Influence of scalar fields on the approach to a cosmological singularity},
(Physical Review D {\bf 61}, 023508 , 1999).

\bibitem{capfirst}
Capozziello, S., \textit{Curvature Quintessence}, Int. J. Mod. Phys. D \textbf{11}  (2002) 483.

\bibitem{capozzielloreview}
Capozziello, S., De Laurentis, M., \textit{Extended Theories of Gravity}, Physics Reports 509 (2011) 167.

\bibitem{odintsovreview}
Nojiri, S., Odintsov, S.D., \textit{Unified cosmic history in modified gravity: from F(R) theory to Lorentz non-invariant models }, Physics Reports 505 (2011) 59.


\bibitem{odintsov}
S. Nojiri, S. D. Odintsov, \textit{Introduction to Modified Gravity and Gravitational Alternative for Dark Energy}, Int. J. Geom. Methods Mod. Phys. 04, 115 (2007)

\bibitem{odintsov2}
Nojiri, S. and Odintsov, S. D. (2008). \textit{Dark energy, inflation and dark matter
from modified f(r) gravity}, arXiv:0807.0685

\bibitem{guo}
J.Q. Guo, D. Wang, and A. V. Frolov, Phys. Rev. D 90, 024017 (2014).

\bibitem{bar1}
J. D. Barrow and S. Cotsakis, Phys. Lett. B, 232, 172-176 (1989).

\bibitem{bar2}
J. D. Barrow and H. Sirousse-Zia,  Phys. Rev. D, 39, 2187-92 (1989).

\bibitem{bar3}
J. D. Barrow and S. Cotsakis, Phys. Lett. B, 214, 515-518 (1988).

\bibitem{del2}
N. Deruelle, Nucl. Phys. B327 253 (1989)

\bibitem{staro1}
Starobinsky, A. A. (1980). \textit{A new type of isotropic cosmological models without singularity}. Physics Letters B \textbf{91} 99-102.

\bibitem{planck}
P. A. R. Ade et al. [Planck Collaboration] ,\textit{Planck 2013 results. XXII. Constraints on inflation}, arXiv:1303.5082 [astro-ph.CO]

\bibitem{gravitation}
Misner, C. W., K. S. Thorne, and J. A. Wheeler, 1973,\textit{ Gravitation} (Freeman Press, San Francisco).

\bibitem{montani-kirillov}
G. Montani, A. A. Kirillov, \textit{Origin of a classical space in quantum inhomogeneous models}, (Zh. Éksp. Teor. Fiz. {\bf 66}, No. 7, 449-453 ,1997).

\bibitem{imponente-montani}
G. P. Imponente, G. Montani, \textit{On the covariance of the mixmaster
chaoticity}, (Physical Review D {\bf 63}, p. 103501, 2001).



\bibitem{ADM}
R. Arnowitt, S. Deser, C. W. Misner, \textit{Canonical variables for general relativity}, (Physical Review \textbf{117}, 6, pp. 1595-1602, 1959).


\bibitem{pesin}
Pesin Ya B (1977) UMN (Russian Mathematical Surveys), \textit{Lyapunov Characteristic
Numbers and Smooth Ergodic Theory} 32, n.4 55-112


\bibitem{arnold}
V. I. Arnold,\textit{ Mathematical Methods of Classical Mechanics}, Springer-Verlag (1989)

\bibitem{MCthesis}
Chitr\'e D. M. (1972) Ph.D. Thesis, University of Maryland

\bibitem{kirillov1997}
A. A. Kirillov, G. Montani (1997) Phys. Rev. D, \textbf{56}, n.10, 6225.

\bibitem{general}
I. M. Khalatnikov and E. M. Lifshitz, Phys. Rev. Lett. \textbf{24}, 76 (1970).

\bibitem{indicikasner}
T. Damour, O. M. Lecian, \textit{Statistical Properties of Cosmological Billiards}, Phys. Rev. D, {\bf 83} 044038, (2011).

\bibitem{segnoq}
  S.~Capozziello and S.~Vignolo,
  \textit{The Cauchy problem for metric-affine f(R)-gravity in presence of perfect-fluid matter},
  Class.\ Quant.\ Grav.\  {\bf 26} (2009) 175013

\bibitem{boj}
Bojowald, M., Date, G. and Hossain, G. M. (2004). \textit{The Bianchi IX model in
loop quantum cosmology}, Class.Quant.Grav. 21, pp. 35413570

\bibitem{segnoq}
  S.~Capozziello and S.~Vignolo,
  \textit{The Cauchy problem for metric-affine f(R)-gravity in presence of perfect-fluid matter},
  Class.\ Quant.\ Grav.\  {\bf 26} (2009) 175013


\bibitem{M95}
G.Montani \textit{On the General Behaviour of the Universe Near Cosmological Singularity}, Classical and Quantum Gravity, 12, 2503, (1995).

\bibitem{K93}
Kirillov, A. A.  \textit{On the question of the characteristics of the spatial
distribution of metric inhomogeneities in a general solution to einstein
equations in the vicinity of a cosmological singularity}, Soviet Physics JETP
76, p. 355., (1993).

\bibitem{K87}
Kirillov, A. A. and Kochnev, A. A. (1987). \textit{Cellular structure of space in
the vicinity of a time singularity in the einstein equations}, Pis ma Zhurnal
Eksperimental noi i Teoreticheskoi Fiziki 46, pp. 345?348.

\bibitem{LK63}
E.M. Lifshitz, I.M. Khalatnikov, \textit{Investigations in relativistic cosmology}, Adv. Physics, 12(46), 185-249 (1963

\end{thebibliography}
\end{document}